# Strong pair correlation in small metallic nanoclusters: the energy spectrum

Yurii N. Ovchinnikov[a,b] and Vladimir Z. Kresin[c8]

a) L.D.Landau Institute for Theoretical Physics, Russian Academy of Sciences, 117334, Moscow, Russia

b) Max-Planck Institute for Physics of Complex Systems, Dresden, D-01187, Germany

c) Lawrence Berkeley Laboratory, University of California at Berkeley, CA 94720

Abstract

The electronic shell structure in small metallic nanoclusters leads to high level degeneracy, which is strongly beneficial for the appearance of pair correlation. This results in a high value of $T_c$ as well as in the appearance of a superconducting gap which causes a strong modification of the energy spectrum. The electronic energy spectrum becomes strongly temperature dependent. Consequently, specific experiments to demonstrate the presence of pair correlation can be proposed.

This paper is concerned with the superconducting state of small metallic nanoclusters ($N \cong 10^2 - 10^3$, where N is the number of the delocalized electrons). The appearance of pair correlation in such clusters was described by us in [1]. In this paper we focus on the nanocluster energy spectrum. It will be shown below that the energy gap parameter drastically affects the spectrum and gives it a strong temperature dependence. This fact allows us to propose a specific experiment to detect the presence of pair correlation.

The shell structure of electronic states in clusters, similar to that in nuclei and atoms, was discovered in [2], see, e.g., the reviews [3,4]. A remarkable feature of many metallic nanoclusters is that their shape, and consequently the energy spectrum strongly depends on the number of delocalized electrons N. "Magic" clusters which contain completely filled electronic shells are characterized by a spherical shape. If the highest occupied shell (HOS) is not completely full, the cluster undergoes a Jahn-Teller distortion, so that its shape becomes ellipsoidal.

Because of the shell structure, the cluster energy spectrum is very different from that expected in a model of equally-spaced energy levels.. For "magic" clusters this leads to a high degeneracy of the HOS. For example, for nanoparticles with N=168 (in this case the orbital momentum of the HOS is L=7) the degeneracy is g=2(2L+1)=30. Such high degeneracy is favorable for pairing. The pairing picture is similar to that in nuclei [5], see the review [6]. The importance of shell structure for pairing in nanoclusters was indicated in [7], and especially in [8].

If the shell is slightly incomplete, it is still realistic for it to have a relatively small level splitting, and the impact of pairing remains strong. Remarkably, pairing in nanoclusters leads to the possibility to observe a superconducting state with $T_C$ much higher than that in bulk samples. Qualitatively, such high values of $T_c$ are due to the high degeneracy of the HOS, i.e., there is a sharp peak in the density of states at the Fermi level; this is similar to the picture introduced in [9].

Pairing is caused by the electron-vibrational interaction. The equation for the pairing order parameter has the form [1]:

$$\Delta(\omega_n)Z = \frac{\lambda T}{W}\sum_{\omega_{n'}}\sum_s \frac{\tilde{\Omega}^2}{\tilde{\Omega}^2 + (\omega_n - \omega_{n'})^2} \bullet \frac{\Delta(\omega_{n'})}{\omega_{n'}^2 + \Delta^2(\omega_{n'}) + \xi_s^2} \qquad (1)$$

Here $\omega_n = (2n+1)\pi T$; we employ the method of thermodynamic Green's functions (see, e.g., [10]), $\tilde{\Omega}$ is the characteristic vibrational frequency, $\xi_s = E_s - \mu$ is the electronic energy referred to the chemical potential, the index "s" labels different energy levels, V is the cluster volume, λ=ην is the bulk coupling constant [11], ν is the density of states, $\eta = \langle I \rangle^2 / M\tilde{\Omega}^2$ is the Hopfield parameter, <I> is the electron-ion matrix element averaged over the states involved in the pairing, Z is the renormalization function (we shall not write out the explicit expression for Z).

The values of $T_c$ for several clusters, e.g., for In, Nb, Zn, were calculated in our paper [1]. Here we consider $T_c$ for Ga and Cd nanoclusters. Indeed, both types of metallic clusters have been observed to display clear shell structure (see, e.g., [12],[13]). Subsequently, we focus on the gap parameter and its temperature dependence; this problem was not discussed in [1]. The evaluation of the spectrum is interesting for its own sake, and, in addition, will allow us to propose an interesting experiment (see below).

Eq. (1) can be written in the following dimensionless form:

$$\phi(x_n) = \delta \sum_{n'} K(x_n, x_{n'})\phi(x_{n'}) \; ; \; n,n' > 0 \qquad (2)$$

Here

$$K(x_n, y_{n'}) = \tilde{\lambda}\left(f^+ + f^- - 4x_{n'}^2 \delta_{nn'} f^+ f^-\right) \sum_s \left[x_{n'}^2 + \chi_s^2 + \phi^2(x_{n'})\right]^{-1}$$

$$f^{\pm} = \left[1 + (x_n \pm x_{n'})^2\right]^{-1} \; ; \; \tilde{\lambda} = \lambda / 2\pi\tilde{\Omega}\nu V \qquad (3)$$

$$x_n = \omega_n \tilde{\Omega}^{-1}; \quad \phi(x_n) = \Delta(x_n)\tilde{\Omega}^{-1}; \quad \delta = 2\pi T\tilde{\Omega}; \quad \chi_s = \xi_s \tilde{\Omega}^{-1}$$

Eqs.(1)-(3) are valid for neutral clusters as well as for ions. Note that for neutral clusters $\tilde{\lambda}$ can be written in the form:

$$\tilde{\lambda} = \lambda \varepsilon_F (2\pi N)^{-1}; \quad \varepsilon_F = E_F \tilde{\Omega}^{-1}.$$

For "magic" clusters Eq. (3) contains a summation over different complete shells, so that $2\sum_s -> \sum_j g_j$, where $g_j$ is the degeneracy of the $j^{th}$ shell. If the shell is incomplete, the label "s" corresponds to the projection of the angular momentum. The position of the chemical potential is determined by conservation of the total number of electrons which can be expressed by the relation

$$N = 2T \sum_n \sum_s G(\omega_n, s) \exp(i\omega_n \tau)\big|_{\tau \to 0} \qquad (4)$$

where the thermodynamic Green's function $G(\omega_n, s)$ is

$$G(\omega_n, s) = -(i\omega_n + \xi_s)\left[\omega_n^2 + \Delta^2(\omega_n) + \xi_s^2\right]^{-1} \qquad (5)$$

At $T=T_c$ one should put $\Phi=0$ in expression (3) (cf. [1]), so that $T_c$ can be calculated from the equation

$$\text{Det}\,|\,1-\delta K(x_n, y_{n'})\,| = 0 \qquad (6)$$

At T=0K the summation over $n$ in Eq.(2) can be replaced by integration. Eq. (4) can be written in the form

$$N = 2\sum_s [1 + \exp(-E_s/T)]^{-1} u_s +$$
$$+ [1 + \exp(E_s/T)]^{-1} v_s \qquad (7)$$

$u_s, v_s = 0.5(1 \mp \xi_s/E_s); \quad E_s = [\xi_s^2 + (E_0^s)^2]^{1/2}; E_0^s$ is determined by the equation

$E_0^s = \Delta\left(i\sqrt{\xi_s^2 + (E_0^s)^2}\right)$, or [see Eq.(3)] $\varepsilon_0^s = \Phi[i((\varepsilon_0^s)^2 + \chi_s^2)^{1/2}]$.

An analysis of Eq. (2) at T=0K allows to determine the order parameter $\Delta(\omega)$, which enters the expression for the thermodynamic Green's function. As is known, the retarded Green's function, whose poles correspond to the energy spectrum, is the analytical continuation of G($\omega$, s ).

Based on Eqs. (6,7), one can calculate the value of $T_c$. Eqs. (2) (with the replacement $\delta\sum_n -> \int_0^\infty dx'$ ) and (7) allow us to evaluate the gap parameter at T=0K. In addition, based on the general equation (1), one can investigate the temperature dependence of the gap. One can see directly from Eqs.(1)-(3) that the values of these quantities for specific clusters are determined by the following parameters: $\tilde{\Omega}$, N, $\lambda$, $E_F$, $\xi_s$. These are known from experimental measurements or from calculations.

As an example, we consider clusters with N=168. As was mentioned above, this choice is determined by large value of the angular momentum of the complete shell (*L*=7), and therefore by the high degeneracy. It is also very important that for this N the energy spacing between the HOS and the lowest unoccupied shell (LUS) is relatively small. Both of these factors are favorable for pairing.

As an example, consider the $Ga_{56}$ clusters (each Ga atom has 3 valence electrons). Using the method described above, we obtain for the critical temperature the value $T_c \approx 140K$. This greatly exceeds that for the bulk, where $T_c^b \approx 1.1K$. The high shell degeneracy is the crucial factor leading to such an increase in $T_c$. We have used the following parameter values: $\tilde{\Omega} = 325K$, $N=168$, $\lambda_b \approx 0.4$, $E_F = 10.4$ eV, $E_H \approx 12.6$ eV.

For Cd nanoclusters ($\tilde{\Omega}=209K$, $N=168$, $\lambda_b \approx 0.38$, $E_F=7.47$ eV, $E_H \approx 9$ eV ) we obtain $T_c \approx 73.5K$ (bulk $T_c^b \approx 0.56K$).

Clusters with partially unoccupied shells undergo a Jahn-Teller shape distortion which splits the degenerate level. On the other hand, removal of electrons from the HOS strongly affects the position of the chemical potential, and this factor turns out to be favorable for pairing. The best scenario would correspond to nanoclusters with slightly incomplete shells (e.g., with $N=166$) and small shape deviations from sphericity. In this case one expect weak level splitting. For example, the HOS becomes a set of close levels classified by the projection of their angular momentum *m*. The picture of splitting is similar to that in atomic nuclei (cf.[14]). To calculate the magnitude of the splitting, one can use the following expression[15]:

$$\delta E_L^m = -2E_L^{(0)} A\{-1 + 3(2L+1)^{-1}[\frac{(L+1)^2 - m^2}{2L+3} + \frac{L^2 - m^2}{2L-1}]\} \qquad (8)$$

where A is the deformation parameter. An explicit expression for the deformation parameter can be found [16] by minimizing the total energy $\delta E = \delta E_{el.} + \delta E_{def}$, where $\delta E_{el}$ is described by Eq. (8) and $\delta E_{def} = 3A^2 V(c_{11} - c_{12})$, with $c_{11}$ and $c_{12}$ the elastic constants (see, e.g., [17]), and V the volume. For N=166 the highest occupied states correspond to L=7, $|m| \leq 7$, and we obtain $A = 0.55 E_H/V(c_{11} - c_{12})$.

With the use of Eqs. (6)-(8) one can calculate Tc for the case of a slightly incomplete shell. For example, for Cd clusters with N=166 we obtain Tc≈90K. For analogous Zn clusters we have Tc≈120 K. A detailed calculation will be described elsewhere.

Let us now turn to the evaluation of the gap parameter and the energy spectrum. As was mentioned above, the gap parameter is described by Eq. (2) with the sum over *n* replaced by integration over *x*. The solution can be sought in the form

$$\phi(x) = \beta_0 (1 + \alpha x^2)^{-1} \qquad (9)$$

The solution is determined by the parameters $\alpha$, $\beta$, and $\mu$ which can be calculated by an iterative method and with the expression (9) as a trial function.

The values of α, β₀ and μ can be obtained by minimization of the quantity <Φ₂-Φ₁>/<Φ₁>; Φ₁ and Φ₂ correspond to the first and second iterations. As a result, we obtain for Zn clusters (N=168)  α=6.8×10⁻², β₀=0.9.  For analogous Ga clusters we obtain α=7×10⁻², β₀=0.7.

Note also that in the absence of pair correlation the smallest excitation energy $\Delta E_{min;0}$ (HOS-LUS interval) at T=0K is equal to $\Delta E = E_L - E_H$ (L≡ LUS, H≡HOS). The chemical potential is located between HOS and LUS, and its position is described by the parameter $\tilde{\mu}$ such that $\mu = E_H + \tilde{\mu}(E_L - E_H)$. In the absence of pairing $\tilde{\mu} = 0.5$ at T=0K.

Pairing has a strong effect on the spectrum ($\Delta E_{min;0} \rightarrow \Delta E_{min;p}$; the label"p" stands for pairing). The value of $\Delta E_{min;p}$ is determined by the relation

$$\Delta E_{min;p} = U_L + U_H \qquad (10)$$

Here

$$U_L = \left[\tilde{\Omega}^2 (\varepsilon_0^L)^2 + \tilde{\mu}^2 (E_L - E_H)^2\right]^{1/2}$$

$$U_H = \left[\tilde{\Omega}^2 (\varepsilon_0^H)^2 + (1-\tilde{\mu})^2 (E_L - E_H)^2\right]^{1/2} \qquad (10')$$

$\varepsilon_0^L$ and $\varepsilon_0^H$ are the gap parameters for the H and L states [see Eqs. (2),(3)].

An analysis based on Eqs. (2),(4) and the values of α, β (see above) leads to the following results for Zn (N=168): $\varepsilon_0^H = 1.1$; $\varepsilon_0^L = 1.55$; $\tilde{\mu}=0.63$. In the absence of pair correlation the square-well model gives $\Delta E_{min;0} \approx 70$ meV. Pairing leads to a noticeable increase in the magnitude of $\Delta E$. In accordance with Eqs. (7),(8), we obtain $\Delta E_{min;p} \approx 95$ meV, so that the value $\Delta E_{min;p}$ is noticeably larger than $\Delta E_{min;0}$. A similar influence of pairing on the excitation energy is found for clusters of other metals such as Ga, Cd. For example, for Ga clusters we obtain $\Delta E_{min;0} \approx 103$ meV, $\Delta E_{min;p} \approx 120$ meV. For Cd clusters $\Delta E_{min;0} \approx 74$ meV, $\Delta E_{min;p} \approx 84.5$ meV. A detailed calculations for other metallic nanoclusters will be described elsewhere.

The effect of pairing on the spectrum is much stronger for clusters with slightly incomplete shells, such as those, e.g., with N=166. As mentioned above, for some clusters of this kind one may expect to see only a small deviation from sphericity and consequently a small degree of energy level splitting. Because the uppermost level of the set formed by the splitting of the HOS is not fully occupied, the absorption edge $\Delta E_{min;0}$ is not large. For example, an estimate based on Eq. (8) leads to a value of ~ 7 meV for the Cd cluster (N=166). Here pairing will lead to a drastic change in the spectrum because of the formation of an energy gap. For example, a calculation of the energy gap for such Cd clusters leads to a threshold value of $\Delta E_{min;p} \approx 41$ meV. Such a significant effect of pairing can be detected experimentally (see below).

Eq. (2) also can be used to evaluate the energy gap parameter near $T_c$. In this region $\Phi_n \ll x_n$, so that an additional term $\propto \Phi_n^2$ should be kept. Calculation leads to the following Ginzburg-Landau expression for the thermodynamic potential:

$$\Omega_s = a\left[-\tau\beta^2 + (2C)^{-1}\beta^4\right] \qquad (11)$$

Here $\tau = 1-T/T_c$, and hence $\beta^2 = C\tau$. For example, after long but straightforward calculations one obtains the following parameter values for N=168: a= 0.6 eV, C=2.7 (for Ga);
a=0.2 eV, C= 1.3 (for Cd); a= 0.76 eV, C=2.9 (for Zn).

Based on Eq. (11), it is possible to estimate the role of fluctuations (cf.[18, 19]). It is worth noting that the large values of $T_c$ and the gap parameter lead to a relatively small coherence length that is comparable with the cluster size; the situation is similar to that in the high $T_c$ cuprates. A straightforward calculation shows that the broadening of the transition is on the order of $\delta T_c/T_c \approx 5\text{-}9\%$. A width of this magnitude noticeably exceeds that of bulk superconductors, but is still relatively small.

The phenomenon of pair correlation discussed here is promising for the creation of high $T_c$ tunneling networks. It would probably require special method of growing isolated clusters in a matrix without a strong disturbance of their shapes and spectra (see, e.g., [20]).

Let us discuss the fundamental question of possible manifestations of pair correlations in small nanoclusters and the possibility of their experimental observation. The phenomenon can manifest itself in odd-even effects for cluster spectra and in their magnetic properties. Such an effect has been observed in [21], but for much larger particles
($N \approx 10^4$-$10^5$). The effect described here is caused by shell structure which results in high values of $T_c$.

The following experiment can be proposed. As described above, pairing results in a strong temperature dependence of the excitation spectrum. At $T>T_c$ the minimum excitation energy is given by $\Delta E_{min;0}= E_L$-$E_H$. At $T<T_c$, pairing modifies $\Delta E_{min;}$, and at low temperatures, close to $T=0K$, the excitation energy strongly exceeds that in the region $T>T_c$. This shift is especially dramatic for clusters with slightly unoccupied shells. Such a change in the excitation energy may be observed experimentally and would represent a strong manifestation of pair correlation. By generating beams of isolated metallic clusters at different temperatures (see, e.g., [22]) in combination with mass spectroscopic size selection would allow one to focus on clusters of specific size at various temperatures. A measurement of the energy spectrum, in particular a determination of $\Delta E_{min}$, for example by the photoelectron spectroscopy technique (see, e.g., [23]), would reveal a strong temperature dependence of the spectrum. For example, for Ga clusters (N=168, $T_c \approx 140K$) one should observe a large difference in $\Delta E_{min}$ in the low temperature region near $T=0K$ and for $T>T_c \approx 140K$. For Cd clusters with N=166 a large difference should be observed for spectra in the low temperature region and for $T>T_c \approx 90K$. The use of Ga or Cd nanoclusters for such experiments looks reasonable, because these materials are superconducting and, as mentioned above, the shell structure of their electronic states has been confirmed experimentally. An experiment of this type would be both realistic and informative.

The authors are very grateful to J.Friedel, A.Goldman, and M.Tinkham for fruitful discussions. We are especially indebted to V.V.Kresin for many useful discussions regarding the properties of nanoclusters and their spectroscopy.

The research of YNO was supported by the CRDF under Contract No. RP1-2565-MO-03 and by RFBR (Russia) . The research of VZK was supported by DARPA under  Contract 05U716.